# Advances in Gaseous Photomultipliers

Rachel Chechik and Amos Breskin[*]

*Dept. of Particle Physics, Weizmann Institute of Science; 76100 Rehovot, Israel*

**Abstract**

We review latest progress in gaseous photomultipliers (GPM) combining solid photocathodes and various types of novel electron multipliers. Cascaded gaseous electron multipliers (GEM) coated with CsI photocathodes can efficiently replace UV-sensitive wire chambers for single-photon recording in Cherenkov and other detectors. Other hole-multipliers with patterned electrodes (Micro-Hole and Strip Plates) and improved ion-blocking properties are discussed; these permit reducing considerably photon- and ion-induced secondary effects. Photon detectors with other electron-multiplier techniques are briefly described, among them GPMs based on Micromegas, capillary-plates, Thick-GEMs and resistive Thick GEMs. The two latter techniques, robust and economically produced, are particularly suited for large-area GPM applications, e.g. in RICH. Cascaded hole-multipliers with very high ion-blocking performance permitted the development and the first demonstration of DC-operated visible-sensitive gaseous photomultipliers with bialkali photocathodes and single-photon sensitivity. Recent progress is described in GPMs operated at cryogenic temperatures for rare-event noble-liquid detectors and medical imaging.

Keywords: *gaseous photomultipliers; gas multiplication; photocathodes; micropattern detectors; photon imaging; visible-sensitive photon detectors*

*PACS*: 29.40-n; 29.40.Cs; 29.40.Ka; 85.60.Ha; 85.60.Gz

## 1. Introduction

Gaseous photomultipliers' main advantage is the possibility of conceiving large-area detectors with multiplication factors that permit efficient imaging of light at single-photon levels. Most of the modern Gaseous Photomultipliers (GPM) can operate at high magnetic fields and at photon fluxes exceeding 1MHz/mm$^2$. While most potential applications are in particle- and astroparticle-physics, many other fields could benefit from this technique. GPMs have been employed since a few decades for relativistic-particle identification, measuring Cherenkov light; particularly, they have been playing an important role in single-photon imaging in Ring Imaging Cherenkov (RICH) systems. Details on the different applications and past GPM techniques can be found in the proceedings of RICH Workshops [1] and in recent reviews [2, 3, 4].

The choice of a particular GPM is naturally dictated by the experiment's requests, e.g.: spectral response, sensitivity, stability, operation properties, lifetime, size and compactness, cost etc.

---
[*] Corresponding author. Tel.: +972-8-9342645; fax: +972-8-9342611; e-mail: amos.breskin@weizmann.ac.il

After many years of intensive R&D and massive use of large-area GPMs with "wire chambers" operating with "gaseous photocathodes", mostly TEA (triethylamine) and TMAE (tetrakis-dimethylamine-ethylene) [1, 2], more recent GPM concepts employ CsI UV-sensitive solid-film photocathodes (PC) [5] coupled to wire-chamber electron multipliers [6, 3]. Examples are the GSI-HADES, CERN-COMPASS, JLAB HALL-A, RHIC-STAR and CERN-LHC-ALICE, as reviewed in [3]. In recent years we have seen a considerable progress in the development of other GPM concepts. The R&D efforts have been generally motivated by the necessity to overcome some basic limitations of wire chambers and the possible extension of GPMs' sensitivity from the UV to the visible spectral range [7, 8]. In wire-chamber GPMs the avalanche develops at the wire vicinity, in an "open geometry", at a few mm from the PC; it results in significant photon- and ion-mediated secondary-avalanches formation, limiting the detector's gain and its single-photon detection efficiency and affecting signal timing and photon localization by broadening the charge induced on the readout elements.

The ion-induced secondary-electron emission is particularly important in GPMs with visible-sensitive PCs; their low electron emission threshold seriously limits avalanche gain in DC mode [9]. Another important consequence of avalanche-ion impact on the PC is its permanent damage, limiting the GPM's lifetime [10]. This effect is particularly strong in photosensitive wire chambers, parallel-plate avalanche chambers (PPAC) and in resistive-plate chambers (RPC), in which all avalanche-originated ions are impinging on the photocathode.

The R&D on gaseous photomultipliers, of many groups, has concentrated in recent years on the search for electron multipliers of a "closed geometry", with reduced photon- and ion-feedback probabilities. While most of the works were focused on UV-sensitive detectors, considerable efforts were devoted also to the development of visible-sensitive GPMs [9, 11, 12].

We will briefly overview the state-of-the-art in this field; we will discuss current techniques, applications and trends in gaseous photomultipliers, referring the reader to more extensive literature.

## 2. Cascaded hole multipliers

### 2.1. General

In cascaded gaseous "hole-multipliers" of different structures discussed below (fig.1), the avalanche develops in successive multiplication stages and is confined within the holes. The hole-diameter is varying between a few tens to a few hundreds of micrometers.

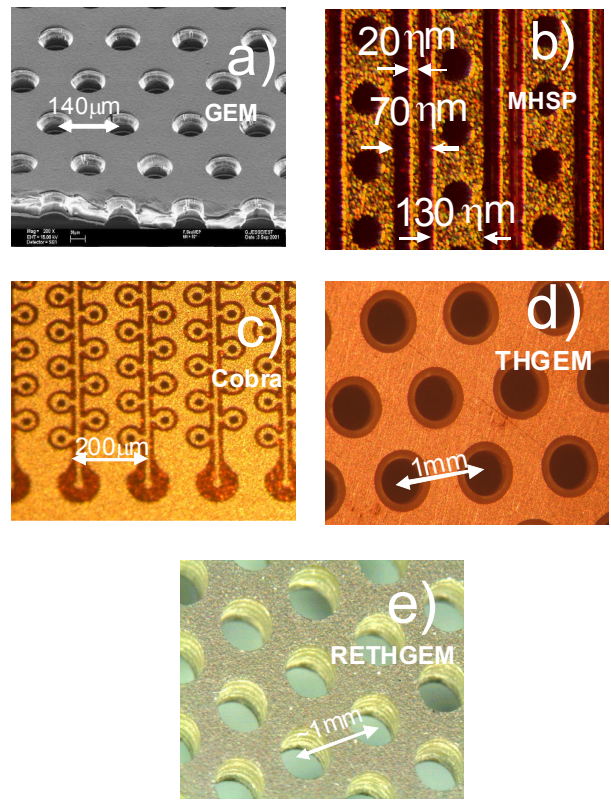

Figure 1. Photographs of different hole-multiplier electrodes described in this review. a) A GEM with 70 micron diameter holes; b) the patterned face of MHSP (the other face is GEM-like) with 50 micron diameter holes; c) the patterned face of the Cobra (the other face is GEM-like) with 50 micron diameter holes; d) a THGEM with 0.4mm diameter holes and 0.1mm etched rim; e) RETHGEM with 0.5mm diameter holes (no rim).

As most of the avalanche-induced secondary photons originate from the last avalanche in the cascade, the preceding elements efficiently screen the PC; this practically prevents photon-feedback effects. Ion-feedback reduction, even in cascaded multipliers, is by far more difficult and challenging; it has been the subject of numerous investigations [13-21]. It is inherently difficult to prevent avalanche ions from back-drifting to the PC while maintaining the full multiplier gain and photoelectron collection and detection efficiencies, because the ions follow the same field lines (though in an opposite direction) as the photoelectrons and avalanche electrons. Efficient methods (discussed below) were recently developed that permit very significantly reducing the Ion Backflow Fraction (IBF), e.g. the fraction of total avalanche-generated ions reaching the PC in a GPM [21].

### 2.2. Cascaded-GEM photomultipliers

Electron multiplication in a Gas Electron Multiplier (GEM) [22], Fig. 1a, occurs in micro-holes (typically 60 microns in diameter) densely etched in a thin double-sided metal-clad insulator (typically 50 microns thick polyimide). Several GEM electrodes can be cascaded and operated with semitransparent [8] or reflective [23] photocathodes; such GPMs reach high multiplication factors (typically $>10^5$), namely single-photoelectron sensitivity. Their operation mechanism and properties with CsI UV-PCs are summarized in [8, 23].

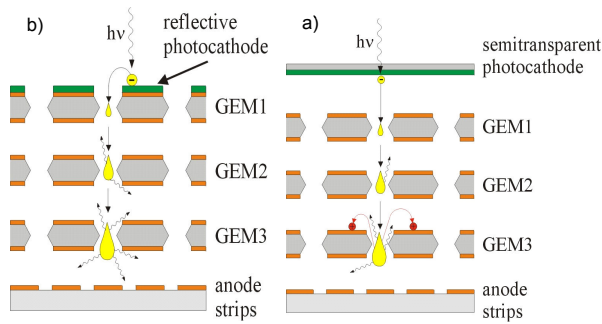

*Figure 2. Schematic views of 3-GEM gaseous photomultipliers with a) semitransparent and b) reflective photocathode.*

Fig. 2 shows 4-GEM GPMs with semitransparent- and reflective- PCs; in the latter, the PC is deposited on top of the first GEM in the cascade [23]. Due to its efficient avalanche-photon screening it reaches gains $>10^6$, in a variety of gases, including $CF_4$ (fig. 3) [24, 25].

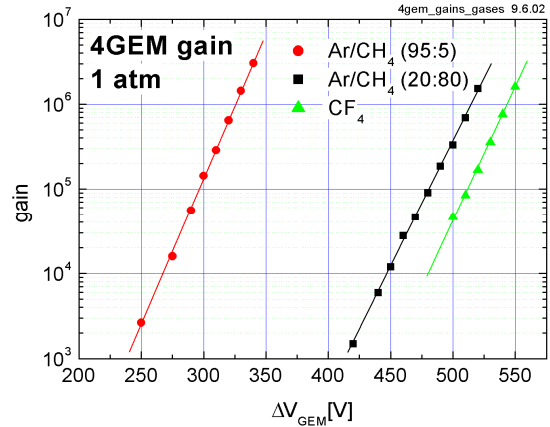

*Fig. 3. Gain vs. voltage across the GEM, of a 4-GEM GPM with a reflective CsI photocathode (similar to the 3-GEM one in Fig. 2b). Gases indicated in the figure.*

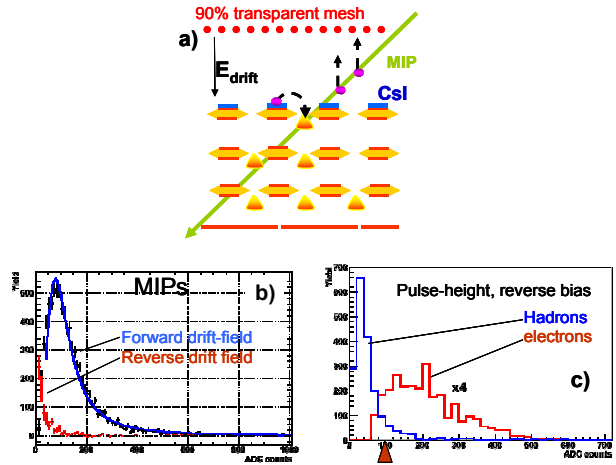

*Figure 4. a) The windowless reflective-CsI 3-GEM GPM of the HBD; the mesh defines a small reversed drift-field above the photocathode, repelling a major fraction of ionization electrons. b) Cluster-amplitude measured at RHIC in $\sqrt{s_{NN}}$=200 GeV p-p collisions, with forward and reversed drift-fields. c) All tracks (hadrons) and electrons measured at RHIC in $\sqrt{s_{NN}}$=200 GeV p-p collisions, with reversed drift-field. Optimizing the threshold at ~channel 100 sets the $e/\pi$ rejection factor to ~85% with ~90% electron detection efficiency.*

The resulting high sensitivity to single photons is due to the high gain of the GEM and its efficient photoelectron collection from the PC, which approaches unity, depending on the gain [8]. The reflective-PC GPM has in addition very low sensitivity to charged-particles background, as discussed in [26]; a small reversed drift-field repels ionization electrons while maintaining good photoelectron collection efficiency (fig. 4).

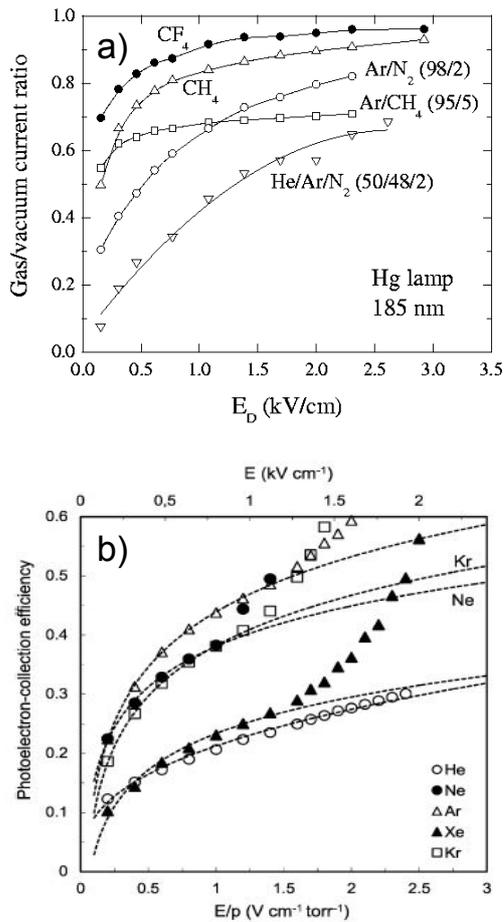

*Figure 5. Measured photoelectron collection efficiency into different gases vs. electric field in: a) molecular gases and gas mixtures; b) noble gases. In the latter the dashed lines are model-calculated; the divergence measured in Ar, Kr, Ne and Xe for E > 1.2kV/cm originates from avalanche-multiplication onset.*

Relativistic-particle rejection factors of ~85% were recently demonstrated in a Hadron-Blind Cherenkov detector [27]. This property is of prime importance in Cherenkov detectors operating under intense background. The effective quantum efficiency (QE) is dictated by the photocathode material and by the photoelectron's extraction efficiency into gas (backscattering). The latter depends on the gas and on the electric field; in Ar/5%$CH_4$, $CH_4$ and $CF_4$ respective extraction efficiencies of the order of 65%, 75%, and 85% were reached (compared to vacuum) at fields of 0.5Kv/cm in 1atm [28] (Fig. 5a); much lower values were measured and estimated in noble gases [29] (Fig. 5b), which may somewhat limit the use of pure noble-gas GPMs. The compact structure of cascaded-GEM GPMs results in short multiplication times; the latter yielded signal pulse-widths in the 10-20 ns range and single-photon time-resolutions <2 ns [30]. The narrow avalanche width permits resolving close-by successive events; the width of the charge induced on the segmented readout anode can be tailored to cope with the readout scheme [31], e.g. by means of a resistive anode in front of the readout circuit [32]. 2D localization resolutions of the order of 100 μm RMS were measured with a 3-GEM detector coupled to a delay-line readout [31]; very good resolutions were obtained with a 3-coordinate readout electrode [33]. Unfortunately, the IBF values in cascaded-GEM GPMs with reflective photocathodes, reached at best values of 10% [15]. Lower values, of 2% were measured with cascaded GEMs coupled to semitransparent photocathodes [13, 14].

### 2.3. Patterned hole-multipliers

With the goal of further reducing the IBF without sacrificing the photoelectron detection efficiency, other hole-multipliers were investigated in different configurations. These have additional strips or other patterns on their surface, with the primary role of deviating avalanche ions. The basic element is a Micro-Hole and Strip Plate (MHSP) [34]; it is a GEM-like hole-electrode with thin anode- and cathode-strips etched on its bottom face (Fig. 1b). Avalanche electrons are multiplied within the hole and additionally on the anode strips. A significant part of the avalanche ions are collected at the cathode strips and on the patterned readout cathode placed below the MHSP; this leads to 4-5 fold smaller IBF in a single-MHSP compared to that of a single-GEM [16]. The MHSP can be used as a stand-alone GPM with a semitransparent- or a reflective-PC, as shown in Fig. 6, for Xe gas scintillation chamber readout [35]; it can be also used as a last multiplying element in a cascade [11], as discussed in the following paragraph.

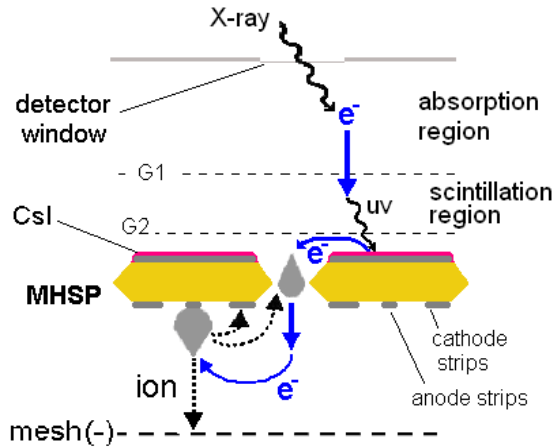

*Figure 6. An x-ray detector combining a Xe windowless gas-scintillation gap and a CsI-coated MHSP GPM.*

The MHSP electrode (Fig. 1b) was further operated in two other modes that could reduce the IBF values in cascaded multipliers. In the Reversed-bias MHSP (R-MHSP) [18, 19], the bias scheme was modified to interchange the roles of anode and cathode strips. The cathode strips can trap a large fraction of back-drifting ions originating from avalanches in subsequent multiplying elements. The operation mechanism and conditions for optimal ion blocking in R-MHSPs, while keeping full photoelectron collection efficiency, are discussed in [19]. A more efficient ion-blocking scheme is offered by the Flipped Reverse-bias MHSP (F-R-MHSP), with the patterned side facing the PC, which traps both its own avalanche ions and ions from subsequent elements [21].

The incorporation of a selection of MHSP, R-MHSP and F-R-MHSP in a multipliers' cascade yielded much lower IBF values, while maintaining full collection of the photoelectrons from the photocathode [19, 21].

The latest variant in the patterned multiplier series is the "Cobra", shown in Fig. 1c; its thin electrodes are curved, surrounding the holes, and the more negatively biased electrodes cover a large fraction of the area, for better ion collection as compared to the F-R-MHSP [36]; its other face is GEM-like. The Cobra multiplier, operated with its patterned surface facing the PC, yielded the best ever measured ion trapping capability [36]. However, in the present electrode's geometry, this came at the expense of a low photoelectron collection efficiency of 20%; it will be presumably improved with better design of the electrode's patterns.

A comparative ion-blocking study was made in 1atm $Ar/5\%CH_4$ with photoelectron extraction field of 0.5kV/cm. Figure 7 shows the IBF values measured with a semitransparent CsI PC in cascaded multipliers with various first-element types, followed by 2 GEMs; the ions generated by the GEMs were trapped by the first element in the cascade. The following IBF values were obtained with single photons at a gain of $10^5$: triple-GEM: 1-2%, R-MHSP followed by a 2-GEM: $3 \times 10^{-3}$; F-R-MHSP followed by a 2-GEM: $2 \times 10^{-3}$; flipped Cobra followed by a 2-GEM: $2 \times 10^{-6}$.

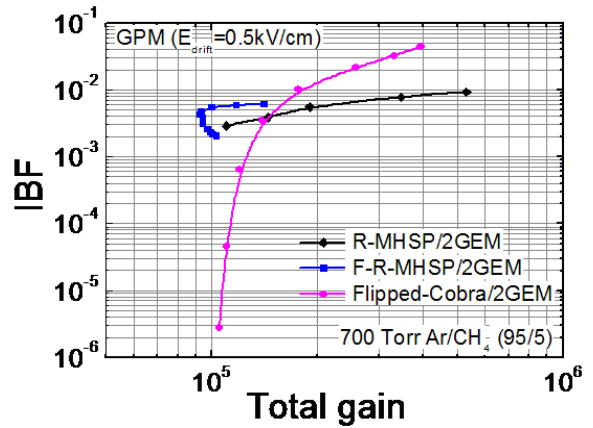

*Figure 7. Measured Ion Backflow Fraction (IBF) to the semitransparent CsI photocathode vs. total charge gain, in a GPM with R-MHSP/2-GEM, F-R-MHSP/2-GEM and F-Cobra/2-GEM multipliers.*

*2.4. Cascaded-photomultipliers with GEMs and patterned hole-multipliers*

We focus here on a recent comparative study of ion blocking in UV-photomultipliers combining CsI photocathodes and cascaded multi-element multipliers; the latter comprise GEMs, MHSPs, R-MHSPs F-R-MHSPs and Cobra [19, 21, 36]. While the MHSP, placed at the end of the cascade, can divert and trap only part of the ions generated within its own avalanche stage, the other types of electrodes can divert ions created in successive multiplying elements (the F-R-MHSP and the flipped Cobra can also block their own ions); therefore the incorporation in the cascade of the different elements, other then GEM, yielded better results. Cascaded-GEM photomultipliers were already discussed above; discussions on the required operation conditions and optimal parameters of the more recent cascaded GPMs can be found in [19, 21].

In all configurations studied, with semitransparent PCs, the drift-field between the photocathode and the first multiplier was kept at 0.5kV/cm (Much lower fields can be applied in Time Projection Chambers (TPC)); this rather high value is necessary for an efficient photoelectron extraction into the gas (low electron backscattering [28]). This constraint indeed restricts the possibility to control the IBF by decreasing the drift field [14]. Therefore, under such conditions, the IBF in semitransparent multi-GEM GPMs reached values of IBF=2% [13, 14]. The operation of a GPM with a reflective PC deposited on the first GEM requests $E_{drift}=0$; the latter results in full photoelectron collection and in efficient rejection of particle-induced ionization electrons [26, 27]. In such conditions, IBF could be reduced at best to levels of ~10% at a gain of $10^5$ [15]. Similarly, in a cascaded multi-GEM/MHSP with reflective PC, the IBF could be reduced to ~2% at effective gains of $10^5$-$10^6$ [16].

The above IBF values are certainly adequate for most applications of CsI-GPMs (CsI exhibits low ion feedback due to low electron-emission probability; however some PC aging occurs due to ion impact, for large accumulated charges [10]), but they are not sufficient for eliminating the considerable ion-feedback observed in visible-sensitive GPMs with K-Cs-Sb PCs [9, 11]; the latter could reach high gains ($>10^5$) only under gated mode [11]. The effective (including backscattering to the PC) secondary electron emission probability γ of bialkali photocathodes was recently measured; its value of 0.03 indicates at the necessity of reaching IBF values of the order of a few times $10^{-4}$ to permit feedback-free stable DC operation at detector gains of $10^5$ [36].

The details of systematic investigations of efficient ion-blocking cascaded-GPM configurations can be found in [19, 21, 36]. Fig. 8 shows three multi-element GPM configurations that yielded low IBF values, sufficient for the DC operation of visible-sensitive GPMs.

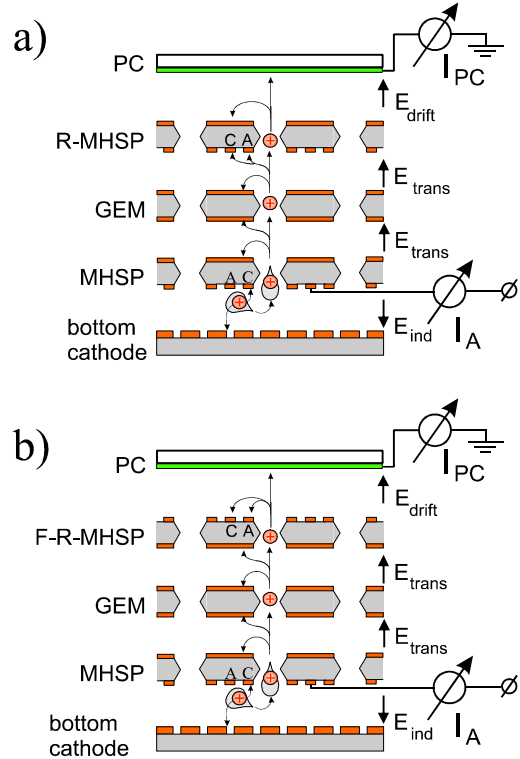

*Figure 8. Schematic views of cascaded gaseous photomultipliers with semitransparent photocathodes coupled to: a) an R-MHSP/GEM/MHSP multiplier and to a F-R-MHSP/GEM/MHSP. Possible ion-paths are shown.*

The IBF values for these configurations are shown in Fig. 9 vs. total gain. The different GPMs yielded the following results at gains of $10^5$: A cascade comprising an R-MHSP followed by a GEM and a MHSP yielded IBF = $9\times10^{-4}$; a cascade of an F-R-MHSP followed by a GEM and a MHSP yielded IBF = $3\times10^{-4}$; both cascades operated at full electron collection efficiencies. The lowest IBF value, of $3\times10^{-6}$ (Fig. 7) was measured with a Cobra followed by 2 GEMs, though with only 20% electron collection efficiency. In all three configurations the low IBF values fulfill our requirement for stable DC operation of visible-sensitive GPMs at gains of $10^5$. An example is presented below.

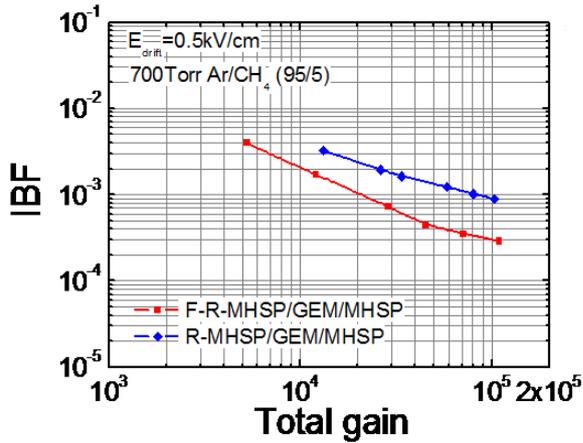

Figure 9. Measured IBF values vs. total gain of the R-MHSP/GEM/MHSP and F-R-MHSP/GEM/MHSP gaseous photomultipliers shown in Fig. 8. The conditions are given in the figure.

*2.5. Thick-GEM and resistive thick-GEM multipliers*

The Thick-GEM (THGEM) [37] has a hole-structure similar to the GEM, but with about 10-fold expanded dimensions (Fig. 1d). It is manufactured economically by mechanically drilling sub-millimeter diameter holes in a thin (generally a fraction of a mm) printed-circuit board (PCB), followed by Cu-etching of the hole's rim. The latter, preventing edge discharges, provides about ten-fold higher gains compared to the "optimized GEM" [38] or the LEM [39]. Like the GEM, two or more elements can be cascaded, to provide very high gains ($>10^6$ with single photoelectrons in a double-THGEM at 1atm Ar/5%CH$_4$ and Ar/30%CO$_2$), thus good single-photon sensitivity (fig. 10a). The same detector yielded gains of $5\times10^3$ and $5\times10^4$ in single- and double-THGEMs with 5.9 keV x-rays, in 1atm Ar/5%CH$_4$.

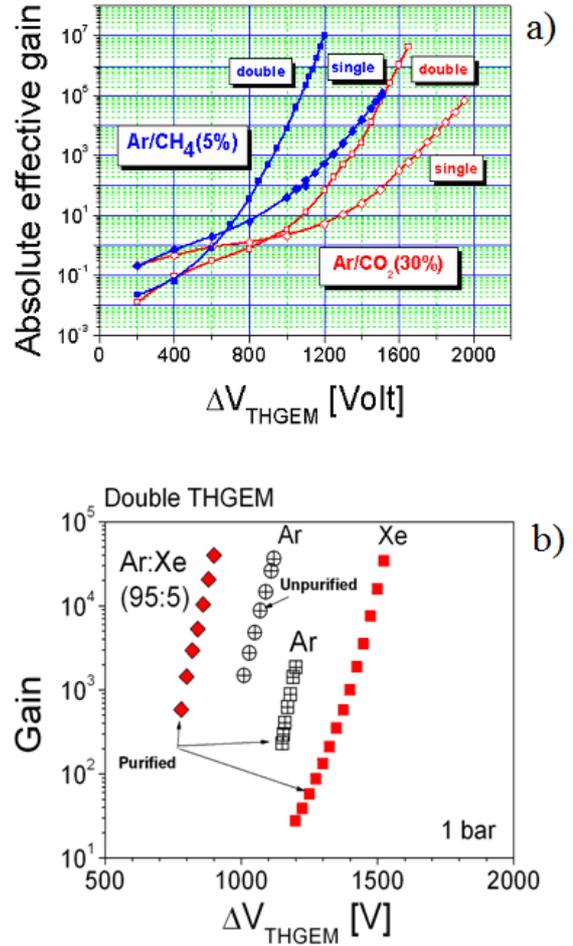

Figure 10. Absolute charge-gain vs. voltage across the THGEM, measured a) with single photons, in single- and double-THGEMs coupled to semi-transparent CsI photocathodes, in Ar/5%CH$_4$ and Ar/30%CO$_2$ and b) with 5.9 keV x-rays, in a double-THGEM, in standard Ar and in getter-purified Ar, Xe and Ar/5%Xe.

Recent investigations of THGEM operation in noble gases and their mixtures yielded gains of $>10^4$ at atmospheric pressure with 5.9 keV x-rays (Fig. 10b) [40]; recent (yet unpublished) investigations in Kr and Ne yielded similar and higher respective gains. The large holes (much larger than electron diffusion) result in good photoelectron collection efficiency into the holes and in a fully efficient THGEM cascading. The efficient cascading requires smaller number of avalanche charges in THGEM- compared to GEM-cascades, for reaching a given total gain. This lead to IBF values of 6% in a semitransparent 2-THGEM GPM in 1atm $Ar/CO_2(70/30)$ at gains of $10^5$ [41]. THGEM-GPMs were investigated in view of their potential application for UV-photon imaging in Cherenkov detectors [41, 42] and in other fields. The simplest configuration, of a double-THGEM with a reflective CsI photocathode deposited on the top surface of the first multiplier, was studied (Fig. 11).

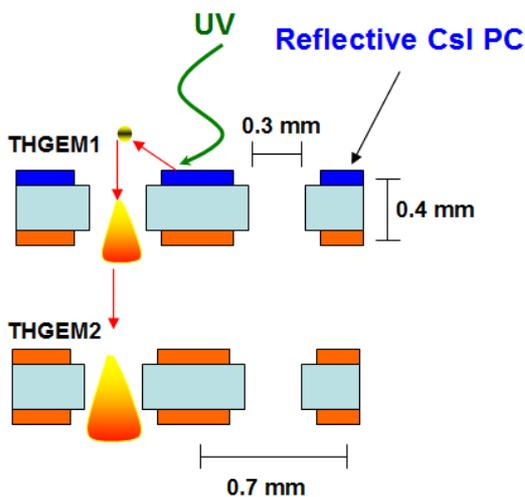

*Figure 11. Schematic view of a double-THGEM with a reflective CsI PC deposited on the top one. Photoelectrons are efficiently focused into THGEM 1 and multiplied in two steps.*

The rate capability measured with single photons, reached fluxes >1MHz/mm$^2$ [41]. Using a very simple readout scheme, localization resolutions of ~0.7mm RMS were demonstrated with this detector [43], which are compatible with most applications of Cherenkov light imaging. The time resolutions were 8ns RMS and 1 ns RMS, with single photons and with 100-photon bursts, respectively (fig 12). Laboratory studies proved the THGEM to be quite robust and very resistant to sparks, compared to GEM. Some initial gain variations, due to charging-up effects (similar to GEM) [44] are under investigations within the CERN-RD51 collaboration project. Details on the THGEM properties, including that of CsI-coated THGEM-GPMs are described in detail in [41-46].

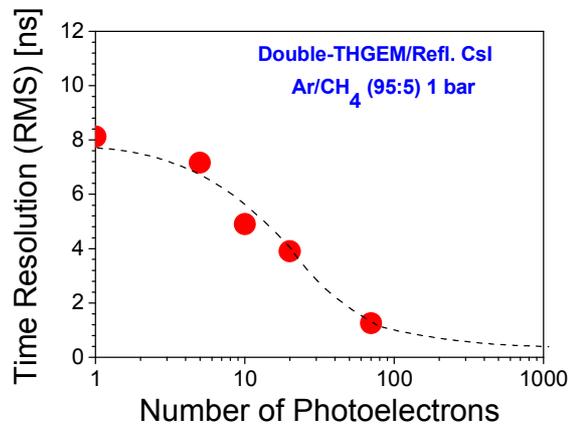

*Figure 12. Time resolution (RMS) vs. number of photoelectrons recorded with a pulsed UV lamp in a reflective double-THGEM GPM with CsI photocathode of figure 11.*

A resistive THGEM (RETHGEM) was recently introduced, in an attempt to conceive a spark-immune multiplier [47]; in the RETHGEM (Fig. 1e) the Cu-clad is replaced by a resistive coating (e.g. resistive Kapton, [48], silk-screen printed surface [49] etc). Like other detectors with resistive surfaces (e.g. RPCs) it has indeed an improved resistance to discharges, but at the expense of lower counting-rate capability - of the order of 10Hz/mm$^2$ for large-area illumination [50] and 100Hz/mm$^2$ for collimated (few mm$^2$) illumination. Gains > $10^5$ were reached in different gases in double-RETHGEM coupled to a CsI PC [51].

## 3. Other GPM concepts

Several GPM concepts have been proposed over recent years; some remained in early conceptual stages while others are under investigations. We will briefly survey some of the techniques; more information can be found in [1].
Before the "GEM-era", GPMs with Microstrip Chambers (MSGC) coated by- or coupled to semitransparent CsI photocathodes were investigated [52, 53], unfortunately displaying poor performance. E.g. gains of ~3 $10^4$ were reached in Ar-$C_4H_{10}$ (90/10) with only few % electron collection efficiency [52]; in a reflective-GPM for a windowless Gas Scintillation Proportional Chamber a gain of 700 was reached in pure Xe [53]. These concepts were abandoned due to their low sensitivity to single photons (low gain and poor photoelectron extraction efficiency) and to operation instabilities.

There have been attempts to develop Micromegas-based photon detectors, with semitransparent photocathodes or with CsI deposited on the mesh electrode [54] (fig 13). The electric-field ratio between the thin parallel-plate multiplication gap and the collection gap was rather favorable for ion blocking (though in TPC operation conditions) [17]. High gains up to $10^6$ were reached in He/isobutane mixtures, where the pulse-height distribution of single photoelectrons had Polya shape; though this should lead to good single-electron detection, the He-based gas mixture would seriously affect the photoelectron extraction from the PC due to backscattering [4]; the UV-cutoff of isobutane will limit the UV spectral response [2]. With reflective PC deposited on the mesh, the optical transparency of the latter limits the effective quantum efficiency. Data on the photoelectron collection efficiency from reflective photocathodes do not yet exist. The R&D efforts continue, triggered by the new possibilities of producing large-area "bulk-Micromegas" detectors [55] and "InGrids" detectors with integrated VLSI readout electronics [56].

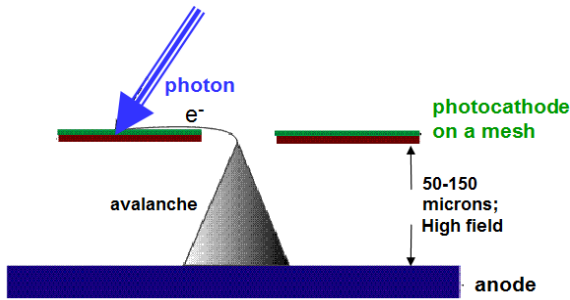

*Figure 13. A Micromegas reflective GPM with CsI-coated mesh.*

The development of GEM-GPMs triggered that of another hole-multiplier, based on glass "capilary-plates" (CP) [57-59]. The latter, with hole-diameters in the hundred microns range, were investigated in single- or cascaded-modes (fig. 14), yielding gains in the order of $10^4$ in Ar/5%CH4 with CsI PCs [58]. Like other hole-multipliers they can operate in pure noble gases; unlike glass-made CPs that are rate-limited due to charging-up effects; low-resistivity hydrogenated CPs can operate at rates reaching $10^5 Hz/mm^2$.

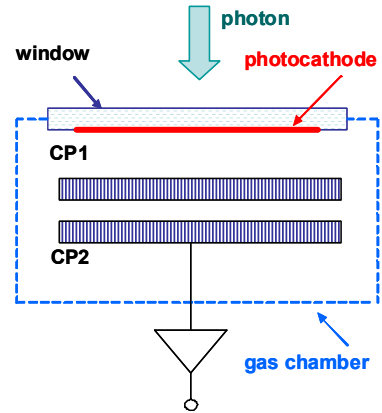

*Figure 14. Double Glass-Capillary GPM with semitransparent photocathode.*

Another interesting solution for ion blocking in GPMs is employing photon-assisted cascaded multipliers (PACEM), as shown in fig. 15 [20]. The idea consists of dividing the amplification chain into two stages, isolated by a reversed electric field, which blocks any charge transport between them (electrons as well as back-drifting ions); the two stages are coupled by scintillation photons produced in the first stage's holes and detected by the second stage's PC (e.g. CsI) and multipliers. Naturally, the concept is applicable only with gases emitting copious photon yields in the UV (e.g. Ar, Kr, Xe, $CF_4$ and their mixtures).

In the search for compact highly-integrated radiation-imaging detectors, new concepts and solutions have been recently proposed. These consist of integrating gaseous multipliers with high-density pixilated readout electronics [60, 61]. The method was demonstrated by high-resolution x-ray imaging, using a GEM coupled to a CMOS chip [60] and in a Micromegas detector coupled to a CMOS chip [62].

In a recent work [63] the UV-photon detector of (Fig. 16) was investigated. It consisted of a semitransparent CsI photocathode coupled to a single-GEM, or a reflective one deposited on its top face; the GEM was coupled to a CMOS VLSI pixel array which acted as a collecting anode and readout electronic circuit.

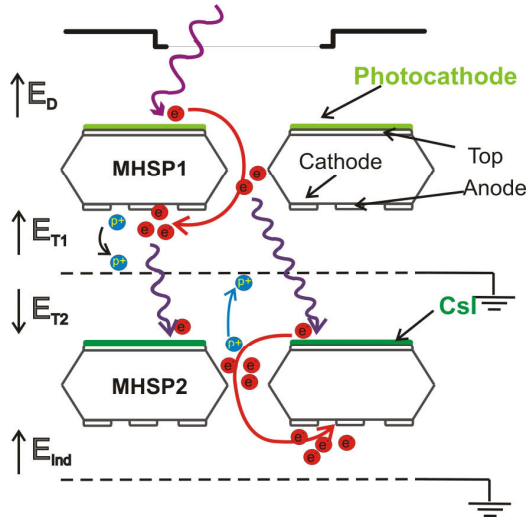

*Figure 15. Operation principle of the Photon-Assisted Cascaded Electron Multiplier (PACEM). Photoelectrons from the GPM's photocathode induce avalanche in the first MHSP element. The electrons from this avalanche are blocked by the grid's reversed voltage. The photons from this avalanche impinge on a second photocathode deposited on the second multiplier, with avalanches developing in successive (not shown) elements. Most ions are blocked at the grid by a reverse field.*

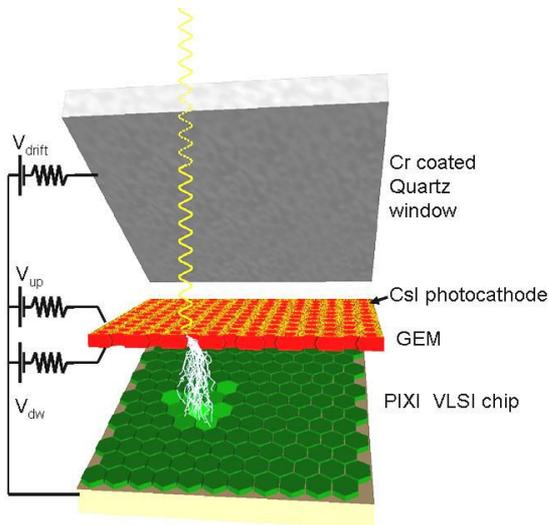

*Figure 16. A single-GEM based high-resolution imaging GPM with a reflective CsI photocathode, coupled to a pad-readout CMOS chip.*

With 50 microns pitch of the GEM and the readout pixels, single-photon record resolutions of 4 microns RMS were reached in a 15x15 mm$^2$ detector [63]. The detector was operated in Ne/DME (50:50), at gains reaching $10^4$. In this configuration and gas mixture, Polya-type charge distributions, well separated from the noise level, were reached, which should permit efficient single-electron detection; the extraction efficiency from CsI into this gas mixture is yet unknown. Like in GEM-GPMs, the ion feedback should be of concern.

## 4. GPMs for the visible spectral range

The dream of detector scientists working in the field of gaseous photomultipliers has always been the extension of their sensitivity to the visible spectral range. It is a very difficult task due to the very high reactivity of visible-sensitive photocathodes (e.g. Bialkali); it results in very short lifetimes in gases with impurity levels in the fraction of ppm range. Therefore, visible-sensitive GPMs can operate, like vacuum photomultipliers, only in sealed containers.

Intensive R&D efforts were made to coat visible-sensitive photocathodes with thin protective films [64, 65]. The idea was to deposit films with good electron transport and escape properties (e.g. Alkali halides); their thickness had to be tuned to provide good protection against gas impurities while maintaining reasonable photoelectron escape. Residual QE values of the order of 6% (4-5 fold lower than of bare PCs) were reached at 330 nm with bialkali photocathodes coated with 20nm thick CsI films; the latter provided good protection against oxygen but not against moisture. References to the works and results are summarized in [7, 66, 67].

First attempts to develop visible-sensitive GPMs with wire-chambers and parallel-grid multipliers CsSb and GaAs-Cs PCs, reached gains below 1000, limited by secondary effects [68]. Cascaded glass capillary-plates, with semitransparent CsSb, AgO-Cs or GaAs-Cs PCs (Fig. 14) provided good screening against photon feedback and some ion blocking on the capillary walls [56]. Made of glass, they are advantageous for sealed GPMs with visible-sensitive PCs, gains close to $10^5$ were reached with such a detector in 1 atm Ar/5%CH$_4$, though with QE values of ~1% at 300-400nm [57]. Somewhat higher QE values, of ~4-6% at 400nm, were reported in [69] with double capillary-plate detectors with CsSb and bialkali PCs operated in 1atm Ar/5%CH$_4$; lower gains, of 1000, resulted from an enhanced ion feedback due the higher QE [12]. In later studies, these authors focused on gas optimization for capillary plates [70]; He-based mixtures (1 atm He/5%isobutane or He/5%Xe) were investigated, with ethylferrocene vapors additives that reduced ion-induced electron emission. Gains above $10^3$ were reached with a single capillary plates plus a parallel-plate multiplier. In attempts to reduce the ion feedback, a "hybrid GPM" was developed [71], with photon multiplication only; photoelectrons from a CsSb PC traverse a screening capillary plate and induce UV-electroluminescence (without charge multiplication) in a parallel-plate structure; the UV photons were detected with a CsI/MWPC at a gain of $10^5$. QE values of 6% were reached in 1 atm CH$_4$ [71].

GPMs with visible-sensitive bialkali PCs and cascaded hole-multipliers with GEM and MHSP-based elements have been intensively investigated over the past few years; we shall point at the highlights of the results, while details are provided elsewhere [8, 9, 11, 30, 72].

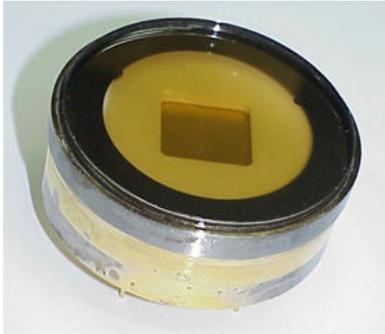

*Figure 17. A photograph of a sealed 3-GEM gaseous photomultiplier with semitransparent K-Cs-Sb photocathode (yellowish color).*

The first significant outcome was the success in keeping semi-transparent bialkali PCs coupled to standard Kapton-made GEMs within indium-sealed detector packages (fig. 17) filled with Ar/5%CH$_4$ for a month period, reaching QE values of the order of 13% at a wavelength of 435nm. In those devices the considerable ion-feedback at the PC, already seen at a gain of couple of hundreds, limited the gain to values <1000 (spark limit) [9]. In such a device the stable high-gain operation was reached by implementing an active ion-gating electrode [15]: a pulsed alternating-bias wire-plane, introduced within the cascaded-GEM structure suppressed the avalanche IBF to the PC by factor of ~$10^4$. This has brought about the next significant landmark: a gated 4-GEM bialkali-GPM (fig. 18a), permitted, for the first time, reaching gains of $10^6$ in the visible spectral range (Fig. 18b).

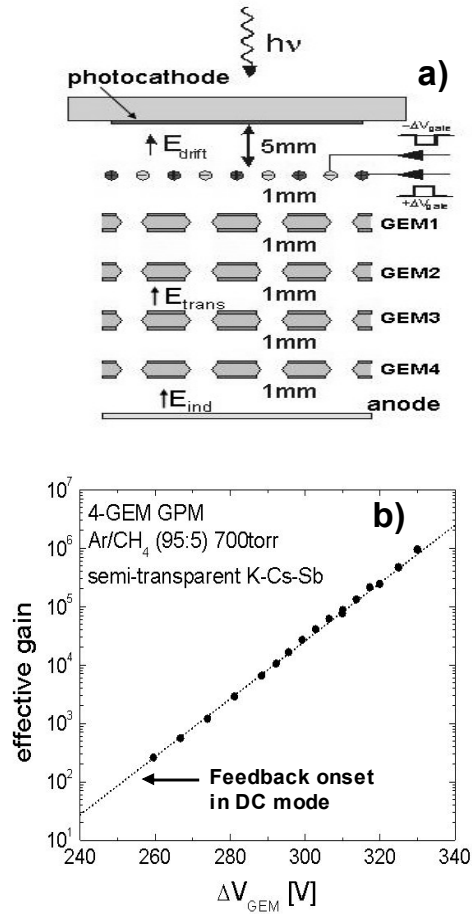

*Figure 18. a) A 4-GEM GPM with pulsed wire-grid ion-gate and a semitransparent K-Cs-Sb photocathode; IBF=$10^{-4}$ b) Charge-gain vs. voltage across the GEM in this GPM; feedback-free gain of $10^6$ was reached with single-photons in gated mode while the ion-feedback onset occurred already at gains of $10^2$ in DC mode.*

However, the gated operation has some drawbacks: the dead-time limits the counting-rate to ~0.1-1 MHz; it is applicable only when a trigger signal is available and it requires extensive screening against the pulser's pick-up noise.

A breakthrough was recently reached in the DC-mode operation of visible-sensitive GPMs [36] comprising a semitransparent Bialkali PC coupled to a cascaded hole-multiplier. The latter was a F-R-MHSP followed by a GEM and an MHSP, described above (Fig. 8b). Stable operation at gains of $10^5$ was reached in DC mode, at atmospheric pressure of $Ar/CH_4$ (95/5) (fig. 19). This validated the hypothesis that an efficient ion blocking (here IBF=$3\times10^{-4}$) permitted, for the first time, operating a visible-sensitive gaseous photomultiplier at such high gains.

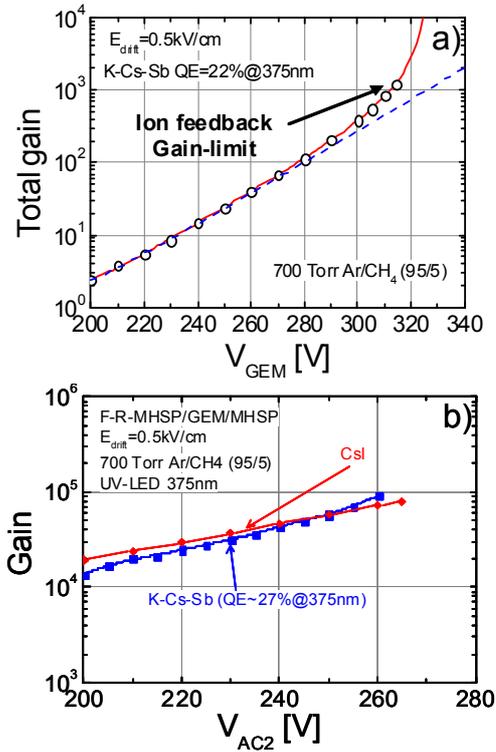

*Figure 19. Charge-gain in DC-mode operation of a visible-sensitive gaseous photomultiplier: a) a 2-GEM with semi-transparent K-Cs-Sb photocathode; gain limit of $<10^3$ is due to ion-feedback; b) a F-R-MHSP/GEM/MHSP (of Fig. 8b) with semitransparent K-Cs-Sb and CsI photocathodes. Note the similarity of the gain curves for both photocathodes, both being free of ion-feedback; measurements were stopped at $10^5$ gain (not a discharge limit).*

Similar results regarding the ion-induced electron emission were reached in K-Cs-Sb, Na-K-Sb and Cs-Sb visible-sensitive photocathodes [36].

The ageing of semitransparent K-Cs-Sb PCs under avalanche-ion impact was recently investigated [11]. Several PCs were studied, at various conditions, showing typically a QE decay of 20% after accumulated charge of 1-2 $\mu C/mm^2$ on the PC, and further 20% QE decay after accumulated charge of 2-4 $\mu C/mm^2$ on the PC. The decay depends on the initial PC surface composition and QE and on the test conditions. The quoted decay-rate is only 4 times faster than that measured for thin, semitransparent CsI PCs [10]. In terms of a photon detector with a bialkali PC, operating at gain=$10^5$ and assuming IBF=$3*10^{-4}$, a 20% QE drop will occur after 46 years of operation at a rate of 5kHz/$mm^2$ photos. The same PC would only survive 5 days in a MWPC-based photon detector under the same operation conditions.

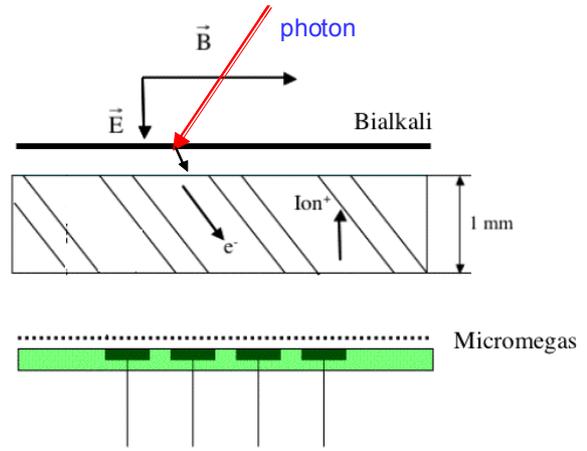

*Figure 20. A GPM concept of a Micromegas multiplier separated from the photocathode by a capillary plate with inclined holes. The latter is designed to transmit photoelectrons under magnetic field (Lorenz-angle) and blocks ions from the Micromegas multiplier.*

There have been an ongoing R&D, in cooperation with industry, of Micromegas-based visible-sensitive sealed GPMs, with a Micromegas multiplier separated from the photocathode by a glass Capillary Plate (CP) with inclined holes (Fig. 20); the latter is designed to block the avalanche ions when operated in a carefully matched magnetic field (Lorenz-angle matched to the channel's inclination) [73]. First results with Bialkali photocathodes operated in different gas mixtures resulted in gains in the few times $10^3$; these are insufficient for single-photon detection [73]. The R&D is in course.

## 5. Cryogenic gaseous photomultipliers

Gaseous photomultipliers could be advantageously applied for recording scintillation light in large-volume rare-event Time Projection Chambers (TPC); these can be high-pressure gaseous detectors, noble-liquid detectors and two-phase (liquid/gas) ones [74, 78], applied in neutrino physics, Dark-Matter searches, double-Beta decay studies etc. They could also record light in noble-liquid Gamma detectors [75]. Important elements in such TPCs are the photon detectors; these operate either in contact with the noble liquid or in the saturated vapor phase above the liquid. They generally detect primary radiation-induced scintillation (e.g. a "start" signal for the TPC); in cryogenic two-phase avalanche detectors they can also detect secondary scintillations induced in the saturated vapor by the extracted (from the liquid) drifting electrons.

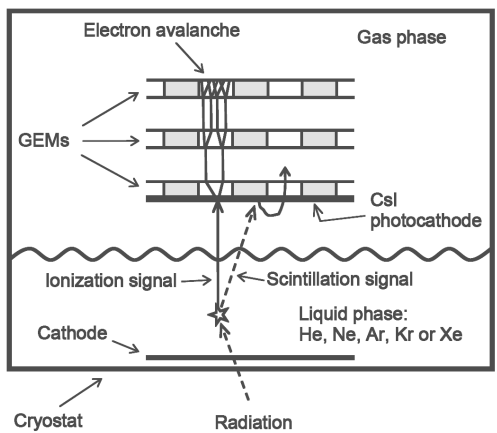

*Fig. 21. Operation principle of the two-phase avalanche detector with a 3-GEM/CsI reflective GPM; both ionization and scintillation signals from the liquid are detected.*

The detection of both scintillation and ionization signals (fig. 21) [80, 81] should provide efficient background rejection in rare-event experiments; in PET applications, the detection of scintillation signals should provide a fast trigger for coincidences between two collinear gamma-quanta.

Various GPMs were investigated in cryogenic conditions in combination with CsI photocathodes by the groups of Peskov et al. and Buzultskov et al.: e.g. single-wire counters, cascaded-GEMs, capillary plates "Optimized GEMs" and THGEMs, as reviewed in [76, 77]; These GPMs operated in a stable way down to $LN_2$ temperatures. It has been known since long that cascaded GEMs permit reaching high gains also in Xe and Ar due the avalanche confinement within the holes [24]; photon-feedback suppression permitted the operation with CsI PCs. This could pave ways towards windowless GPMs, operating in the vapor of two-phase detectors, though with the drawback of considerable photoelectron backscattering losses (QE losses).

There have been numerous works on hole-multipliers operating in cryogenic conditions, without and with CsI photocathodes. E.g. it was proven that cascaded-GEMs could reach gains $>10^4$ at low temperatures and in two-phase mode in Ar and Kr [78, 77, 80]. The stability, QE and gain in cryogenic cascaded-GEMs and capillary plates are discussed in [79]. Recent investigations of a two-phase Ar avalanche detector with a reflective-CsI triple-GEM GPM yielded charge gains of $10^4$ [81]. Fig. 22 illustrates a scintillation signal in this two-phase detector, induced by beta-particles, at a gain of 2500. Both, scintillation and ionization signals are clearly observed. The amplitude of the scintillation signal corresponds here to ~30 photoelectrons.

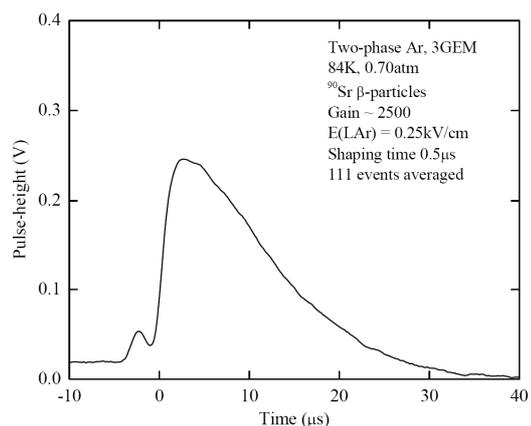

*Figure 22. Scintillation (1st) and ionization (2nd) signals in the 3-GEM/CsI GPM of Fig. 21, in a two-phase LAr detector. Gain 2500; signals induced by Beta-particles.*

Photosensitive THGEMs were investigated at cryogenic temperatures; gains of ~$10^3$ and ~$10^4$ were reached with a double-THGEM operated in 1atm Xe or Ar, detecting alpha-induced scintillations or single photoelectrons, respectively [82]. Recent comparative investigations of a two-phase liquid-Ar detector with a triple-GEM and double-THGEM yielded gains of 6000 for the latter in the saturated vapour phase [77]. Last but not the least is the CsI-coated double-RETHGEM that reached gains close to $10^4$ when placed at 20mm above LAr [83].

## 6. Summary and applications

Gaseous photomultipliers continue playing an important role in many scientific fields. They compete with other technologies when imaging of low light levels (down to single photons) over large detection areas is of a concern, as in Cherenkov detectors. Advanced concepts described in this work permit conceiving compact single-photon imaging detectors. Some can be of relatively low mass, operate at high magnetic fields or/and at cryogenic temperatures, provide single-photon time resolutions in the few-ns range, localization resolutions of a few-hundred microns and rate capabilities in the MHz/mm$^2$ range. Recent advances in avalanche-ion blocking permitted, for the first time, the conception of visible-sensitive gaseous photomultipliers with bialkali photocathodes, operating in DC-mode at high gains. These could pave the way to numerous applications, beyond that of Cherenkov-light imaging.

Out of the GPM concepts reviewed in this work, cascaded-GEM GPMs reached the stage of application. GPMs combining cascaded GEMs with reflective CsI photocathodes are presently tested for the first time in a RICH system; large-area detectors are in operation in a Hadron-Blind Cherenkov detector (HBD) of the PHENIX (RHIC-BNL) relativistic heavy-ion experiment up-grade [27, 84, 85]. The HBD shown in figure 23 exploits the unique properties of GEM-GPMs (e.g. operation in $CF_4$, reduced sensitivity to ionizing radiation) in order to comply to the geometrical and physical constraints at this experiment. About 1 m$^2$ of windowless GPMs, mounted on a 0.6m radius barrel, are coupled in proximity to the $CF_4$-filled radiator volume, and operate in this gas. The GPMs, 24 in total, are 3-GEM cascades with a reflective CsI photocathode (similar to the detector of Figs. 2b, 4a), operate in the reversed drift-field mode. Due to the windowless structure, the HBD system has an unprecedentedly high figure of merit ($N_0$) [2] (design value $N_0$=822cm$^{-1}$ [27]). The signals are recorded on hexagonal readout pads, designed to show the relativistic-electron hits as 3.6cm diameter "blobs" occupying several pads, which are well distinguished from residual hadron-background hits occupying typically a single pad (fig 24).

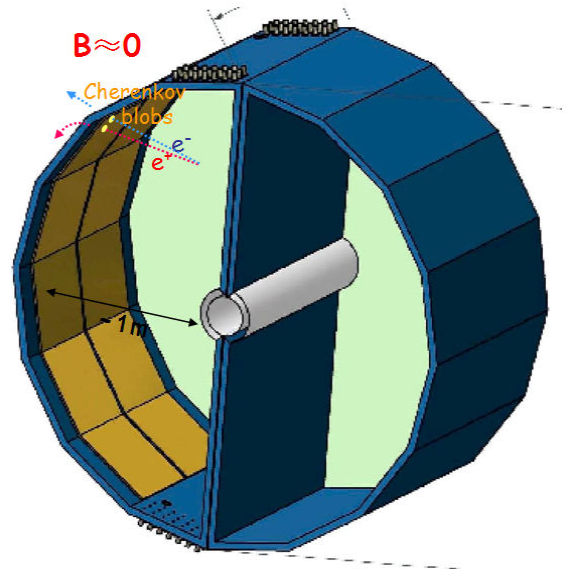

*Figure 23. Schematic view of the hadron-blind detector (HBD) at RHIC-PHENIX. It consists of two containers; the front and back planes were removed for clarity. The HBD fits into a limited volume of zero magnetic field. 24 windowless 3-GEM/CsI GPMs are mounted on the barrel, in proximity to the radiator volume, filled with $CF_4$.*

The entire RICH system was constructed with minimum mass, so as to fit under 3%$X_0$. Engineering runs during 2007 (Au-Au collisions at 200GeV) demonstrated the HBD operation at gas gain of ~5000, with noise pedestal rms equivalent to 0.2e$^-$, and confirmed good relativistic-electron detection efficiency with good separation from hadrons. The use of $CF_4$ as detector gas requires rather high operation voltages on the GEM electrodes; this, in turn, requires great care in the electrodes' selection and handling (i.e. electrode quality control, cleanliness during construction and storage, etc). The particle-induced scintillation light from $CF_4$ was found to be of no limitation to the performance of the HBD in heavy ion reactions [86].

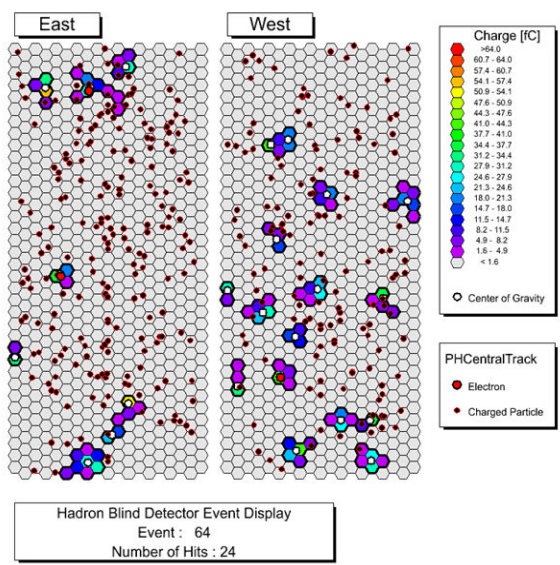

*Figure 24. MC-simulated HBD event, from central Au-Au collision at $\sqrt{s}_{NN}=200$ GeV. Electrons are recorded as broad blobs of 3 pads on average, while background ionization events extend over 1.2 pads on average. $CF_4$ scintillation is not included.*

Recent progress in the operation of cascaded-GEM -THGEM and -RETHGEM GPMs at cryogenic temperatures (that of liquid-Ar and -Xe) and in two-phase detectors, should pave the way towards their potential application in rare-event detectors, e.g. dark-matter, neutrino-scattering and double-beta decay. R&D is in course for these applications; e.g. for the XENON Dark-Matter experiment [87] THGEM electrodes made of low-radioactivity materials (e.g. Cirlex) are being investigated [88]. THGEM-GPMs are under development for recording liquid-Xe scintillations in a Compton Camera developed for a 3-photon medical imager within a collaboration project of Subatech-Nantes and Weizmann Institute [75 grignon].

Both the double-THGEM and double-RETHGEM with reflective CsI photocathodes are under study for RICH-detector upgrades of CERN-COMPASS and -ALICE; methodes for their effective production and characterization are under investigation within the CERN-RD51 collaboration.

## Acknowledgments

The work reviewed here was supported in part by grants from the Israel Science Foundation and the MINERVA Foundation. We acknowledge the assistance of numerous colleagues: V. Peskov, A.Buzulutskov, J.dos Santos, J. Veloso, F. Amaro, J. Maia, J. Miyamoto, M. Cortesi, A. Lyashenko and others. A.B. is the W.P.Reuther Professor of Research in peaceful uses of atomic energy.

## References

[1] Proceedings of the International RICH Workshops and references therein: Nucl. Instr. and Meth. Volumes: A343(1993); A371(1996); A433(1999); A502(2003), A553(2005) and these RICH07 proceedings.
[2] E. Nappi and J. Seguinot, La Revista del Niovo cimento, Vol 28, No. 8-9, 2005 and refs. Therein.
[3] F. Piuz, Nucl. Instr. and Meth. A502(2003)76 and refs. Therein.
[4] A.F.Buzulutskov, Physics of Particles and Nuclei, Vol.39, No.3, pp.424-453; Pleiades Publishing Ltd. 2008. and refs. Therein.
[5] A. Breskin, Nucl. Instr. and Meth. A371(1996)116 and refs. Therein.
[6] V. Dangendorf et al. Nucl. Instr. and Meth. A289(1990) 322.
[7] A. Breskin et al., Nucl. Instr. & Meth. A442(2000)58 and references therein.
[8] R. Chechik et al., Nucl. Instr. & Meth. A502(2003)502 and references therein.
[9] M. Balcerzyk et al, IEEE Trans. Nucl. Sci. NS50(2003)847 and references therein.
[10] B. K. Singh et al., Nucl. Instr. & Meth. A454(2000)364.
[11] A. Breskin et al. Nucl. Instrum. & Meth. A553(2005)46.
[12] T. Francke et al. Nucl. Inst. & Meth. A525(2004)1 and references therein.
[13] A. Breskin et al., Nucl. Instr. & Meth. A478(2002)225.
[14] A. Bondar et al, Nucl. Instr. & Meth. A496(2003)325.
[15] D. Moermann et al. Nucl. Instr. & Meth. A516(2004)315.
[16] J. M. Maia et al., Nucl. Instr. & Meth. A523(2004)334.
[17] P. Colas et al. Nucl. Instr. & Meth. A535(2004)226.
[18] S. Roth et al., Nucl. Instr. & Meth. A535(2004)330.
[19] A. Lyashenko et al. 2006_JINST_1_P10004.
[20] J.F.C.A. Veloso et al. 2006_JINST_1_P08003.
[21] A. Lyashenko et al. 2007_JINST_2_P08004.
[22] F. Sauli, Nucl. Instr. & Meth. A386(1997)531.
[23] D. Mörmann et al., Nucl. Instr. & Meth. A530(2004)258 and references therein.

[24] A. Buzulutskov et al, Nucl. Instrum. & Meth. A443(2000)164 and A442(2000)68.
[25] A. Breskin et al., Nucl. Instr. and Meth. A483(2002)670.
[26] D. Mörmann et al., Nucl. Instr. & Meth. A471(2001)333 and A478(2002)230.
[27] Z. Fraenkel et al., Nucl. Instr. and Meth. A546(2005)466.
[28] A. Breskin et al., Nucl. Instr. and Meth. A483(2002)670 and refs. therein.
[29] L.C.C. Coelho et al., Nucl. Instr. and Meth. A581(2007)190 and refs. therein.
[30] D. Mörmann et al., Nucl. Instr. & Meth. A504 (2003) 93.
[31] G. Guedes et al., Nucl. Instr. & Meth. A497(2003)305 and A513(2003)473.
[32] V. Dangendorf et al., Nucl. Instr. & Meth. A535 (2004) 93.
[33] F. Sauli, Nucl. Instr. & Meth. A553(2005)18.
[34] J. F. C. A. Veloso et al., *Rev. Sci. Instrum.*, 71 (2000) 2371.
[35] E.D.C. Freitas et al., Nucl. Instr. and Meth. A580 (2007) 214.
[36] A. Lyashenko et al., INSTR08, Novosibirsk Feb. 2008. arXiv:0804.4396; to be published in Nucl. Instr. & Meth. A.
[37] R. Chechik et al., Nucl.Instr. and Meth. A535(2004)303.
[38] L. Periale et al., Nucl. Instr. and Meth. A478(2003)377.
[39] P. Jeanneret, *Time Projection Chambers and detection of neutrinos*, PhD thesis, Neuchâtel University, 2001.
[40] R. Alon et al. 2008 JINST 3 P01005
[41] C. Shalem et al., Nucl. Instr. and Meth. A558(2006)475.
[42] R. Chechik et al., Nucl. Instr. and Meth. A 553(2005)35.
[43] M. Cortesi et al., 2007 JINST 2 P09002.
[44] R. Chechik et al., Proc.of the SNIC Symposium on novel detectors, Stanford, Ca-USA, April 2006. http://www.slac.stanford.edu/econf/C0604032/papers/0025.PDF
[45] C. Shalem, MSc Thesis, Weizmann Institute of Science. http://jinst.sissa.it/jinst/theses/2005_JINST_TH_001.pdf
[46] R. Alon, MSc Thesis, Weizmann Institute of Science. http://jinst.sissa.it/jinst/theses/2008_JINST_TH_001.pdf
[47] J.M.Bidault et al., Nucl. Phys B (Proc. Suppl.) 158(2006)199.
[48] R. Oliveira et al., http://arxiv.org/abs/physics/0701154
[49] G. Agocs et al., 2008 JINST 3 P02012.
[50] G.Agocs et al., These RICH07 Proceedings.
[51] A. Di Mauro et al, Nucl. Instr. and Meth. A 581(2007)225
[52] K. Zeitelhack et al., Nucl. Instr. & Meth. A351(1994)585 & A371(1996)585.
[53] C.M.B. Monteiro et al., IEEE Trans. Nucl. Sci. 49(2002)907.
[54] J.Darre et al., Nucl. Instr. and Meth. A449(2000)314.
[55] I. Giomataris et al, Nucl. Instr. & Meth. A 560(2006)405.
[56] V.M. Blanco Carballo et al., Nucl. Instr. and Meth. A576(2007)1.
[57] V. Peskov et al., Nucl. Instr. and Meth.A433(1999) 492.
[58] V. Peskov et al., IEEE Trans. Nucl. Sci. 47(2000)1825.
[59] I. Iacobaeus et al., Nucl. Instr. & Meth.A525(2004)42 and refs. therein.
[60] R. Bellazzini et al., Nucl. Instr. & Meth.A535(2004)477.
[61] J.L. Visschers et al., Nucl. Instr. & Meth.A572(2007)203.
[62] P. Colas et al., Nucl. Instr. & Meth.A535(2004)506.
[63] R. Bellazzini et al., Nucl. Instr. & Meth.A581(2007)246.
[64] V. Peskov et al., Nucl. Instr. & Meth.A348(1994)269.
[65] A. Breskin et al., Appl. Phys. Lett. 69(1996)1008.
[66] E. Shefer et al., J. Appl. Phys. 92 (2002) 4758.
[67] E. Shefer, Ph.D. Thesis, Weizmann.Institute, http://jinst.sissa.it/jinst/theses/2000_JINST_TH_001.pdf
[68] V. Peskov et al. Nucl. Instr. & Meth.A367(1995)347.
[69] V. Peskov et al., IEEE Trans. Nucl. Sci. 47(2000)1825.
[70] V. Biteman et al., Nucl. Instr. & Meth.A471(2001)205.
[71] I. Rodionov et al., Nucl. Instr. & Meth.A478(2002)384.
[72] D. Moermann, Ph.D. Thesis Weizmann Institute, http://jinst.sissa.it/jinst/theses/2005_JINST_TH_004.pdf
[73] J.Va'vra & T. Sumiyoshi, Nucl. Instr. and Meth. A435(2004)334 and A553(2005)76.
[74] B.A. Dolgoshein et al., Physics of Elementary Particles and Atomic Nuclear 4(1973)167 (in Russian).
[75] C.Grignon et al., Nucl. Instr. and Meth.A571(2007)142 and refs. therein.
[76] L. Periale et al., Nucl. Instr. and Meth. A535(2004)517 and references therein.
[77] A. Bondar et al. and references therein, arXiv:0805.2018, Submitted to JINST (May 2008).
[78] A. Buzulutskov et al., IEEE Trans. Nucl. Sci. 50(2003)2491.
[79] L. Periale et al., IEEE Trans. Nucl. Sci. 52(2005)927.
[80] A. Bondar et al., Nucl. Instr. and Meth. A556(2006)273.
[81] A. Bondar et al., Nucl. Instr. and Meth. A581(2007)241.
[82] L. Periale et al., Nucl. Instr. and Meth. A567(2006)381 and A573(2007)302.
[83] V. Peskov et al. IEEE Trans. Nucl. Sci. 54(2007)1784.
[84] A. Koslov et al, Nucl. Instr. and Meth. A523(2004)345-354.
[85] S. Milov et al, J. Phys. G: Nucl. Part. Phys. 34(2007)S701.
[86] B.Azmoun et al, A measurement of scintillation in $CF_4$ using GEM foils and a CsI photocathode, poster presented at the IEEE-NSS symposium, 2007. To be published in the conference records.
[87] E. Aprile et al. Columbia Univ. http://arxiv.org/abs/astro-ph/0207670v1
[88] M.Gai et al., Proc. 23rd Winter Workshop on Nuclear Dynamics, Big Sky, Montana, USA, Feb. 11-18, 2007. arXiv:0706.1106